\documentclass[usenatbib]{mn2e}

\usepackage{graphicx}

\title[The Balmer Edge of the Quasar Big Blue Bump]{A First Close Look at
the Balmer Edge Behavior of the Quasar Big Blue Bump}

\author[Kishimoto, Antonucci \& Blaes]{Makoto
Kishimoto$^{1}$\thanks{E-mail: mk@roe.ac.uk}, Robert Antonucci$^{2}$
and Omer Blaes$^{2}$\\ $^{1}$Institute for Astronomy, University of
Edinburgh, Blackford Hill, Edinburgh EH9 3HJ, UK\\ $^{2}$Physics
Department, University of California, Santa Barbara, CA 93106, USA}

\begin{document}

\date{\today}

\pagerange{\pageref{firstpage}--\pageref{lastpage}} \pubyear{2003}

\maketitle

\label{firstpage}

\begin{abstract}

We have found for the first time a Balmer edge feature in the polarized
flux spectrum of a quasar (Ton 202).  The edge feature is seen as a
discontinuity in the slope, rather than as a discontinuity in the
absolute flux.  Since the polarized flux contains essentially no broad
emission lines, it is considered to arise interior to the broad emission
line region, showing the spectrum with all the emissions outside the
nucleus scraped off and removed.  Therefore, the polarized flux spectrum
is likely to reveal features intrinsic to the Big Blue Bump emission.
In this case, the existence of the Balmer edge feature, seen in
absorption in the shorter wavelength side, indicates that the Big Blue
Bump is indeed thermal and optically thick.

\end{abstract}


\section{Introduction}\label{sec-intro}

Studies of the spectral energy distribution of quasars have shown that,
among the various apparent components present, the most energetically
dominant component is in the optical/UV (e.g. \citealt{Sa89}).  However,
rather surprisingly, the emission mechanism of this component, often
called the Big Blue Bump (BBB), has not been well understood.  Although
the BBB is often assumed to be from optically-thick thermal emission
from an accretion disk, observations are hardly said to be well
described by simple disk models \citep{An88,An99,KB99}.  The current
situation is that various disk models are still trying to accommodate
the main features observed.

Among several serious problems are the continuum slope and apparent lack
of continuum edges. In order to explain the observed optical/near-UV
slope (e.g. $F_{\nu} \propto \nu^{-0.3}$, \citealt{Fr91};
\citealt{Ne87}), the overall disk temperature has to be rather low
\citep{KB99}.  However, naively, at least in such a cool disk, its
atmosphere would show large continuum edges. Apparently, we do not
observe such features at the Lyman edge \citep{AKF89,KKB92}.  This could
possibly be accommodated by sophisticated disc atmosphere models
(e.g. \citealt{Hu00}), where the spectrum integrated over radii are
smeared through the relativistic Doppler shifts and gravitational
redshifts.  In these models, the flux discontinuity at the Lyman limit
is rather well smeared. There is instead a slope change near the Lyman
limit which is accompanied by an emission bump just longward of the
change \citep{Hu00}. Some quasar composite spectra
(e.g. \citealt{Zh97}), as well as some individual spectra (e.g. 3C273,
\citealt{Kr99}), show a slope change near this wavelength.  But the
predicted emission bump is not seen and it appears at the moment that
the model slope change is not in a satisfactory agreement with the
observed ones \citep{Bl01}.

Another problem is the magnitude and direction of polarization.  A
simple disk atmosphere would show the polarization equal to the
classical case of the plane parallel electron scattering atmosphere
which has $P$ of $0-11.7$\% for the inclination angle of $0-90$\degr,
with the direction of polarization (direction of Electric vector
vibration) perpendicular to the disc symmetry axis \citep{Ch60}.  The
observed polarization of the BBB emission is typically $\sim1$\%, and
the position angle of polarization is {\it parallel} to the radio jet
structure, when observations are available, which is presumed to be
parallel to the disc axis.  More recent calculations show that
absorption opacity and general relativistic effects \citep*{LNP90} and
Faraday depolarization with magnetic fields in the disc atmosphere
\citep{AB96,ABI98} can reduce the magnitude of polarization.  The
inconsistency in the direction of polarization will still remain, but
absorption opacity effects might in some cases flip the polarization
direction (first noted by D.I. Nagirner, as described by
\citealt{GS78}), and a rough surface of the discs might lead to the
parallel polarization \citep{CS90}.  Alternatively, the disk radiation
is completely depolarized and then re-polarized slightly in the parallel
direction by surrounding gas.  We also note that there is evidence of
very high polarization shortward of Lyman edge in some objects
(e.g. \citealt{Ko95}).

While the Lyman edge feature could be smeared by the relativistic
Doppler shifts and gravitational redshifts in the disk atmosphere
models, these will be less effective at smearing the Balmer edge
feature. This is simply from a robust prediction of the disk models that
the Balmer edge region of the spectrum originates from farther out in
radius than the Lyman edge region. This should all be true even though
the accretion disc stellar atmosphere models of the Lyman and Balmer
edge features are still plagued by considerable uncertainties (see
e.g. \citealt{Hu00}, \citealt{Hu01} for discussion).  Comptonization may
also smear the Lyman edge \citep{CZ91}, but as this is more important in
high temperature regions, the Balmer edge should be less affected (see
e.g. Fig.8 of \citealt{Hu01}).  In addition, aside from the disk models,
the Lyman edge feature, which is due to a resonant transition, could
rather easily be imprinted by a foreground absorption, but this is less
likely for the Balmer edge.  The Balmer edge should therefore be a key
target wavelength region.  The observation of this wavelength range not
only could test an accretion model, but more fundamentally, could
extract direct evidence for the thermal nature of the BBB emission.
Even this fundamental issue has not been shown directly, and this lack
of direct evidence for the emission nature has been one of the main
sources of the controversy.

The only problem with checking this empirically is that the Balmer edge
wavelength region of the BBB emission is very hard to observe due to the
Balmer emission lines/continuum and FeII emission lines from the Broad
Line Region (BLR) and outer regions (called the 3000\AA\ bump, or the
small blue bump).  However, we can sometimes remove all these unwanted
emissions by taking a polarized flux spectrum.  As mentioned above, many
quasars are found to be polarized at $P \sim 1$\%, and at least in
some cases, emission lines (broad and narrow) are unpolarized - the
polarization is confined to the continuum (\citealt{SS00};
\citealt{An88}, showing the data of Miller and Goodrich). This indicates
that the polarization mechanism resides interior to the BLR in these
cases. Thus, the polarized flux shows the spectrum interior to the BLR,
revealing the underlying behavior at the Balmer edge region.

We have obtained high S/N spectropolarimetry data of quasars with the
Keck telescope, which allow us to investigate this Balmer edge behavior
in detail. We report in this paper that we have indeed found a Balmer
absorption feature in one of the quasars we observed.

\begin{table}
  \caption{Observation log at Keck I}
  \begin{tabular}{lccc}
  \hline
  Name        & redshift & slit PA (\degr) &   Exposure time (min)\\

  \hline
  Ton 202     & 0.366 &  72 & 30 $\times$ 2 \\
  Ton 202     &       & 102 & 30 $\times$ 2 \\
  4C37.43     & 0.371 &  95 & 30 $\times$ 2 \\

  \hline
  \end{tabular}
  \label{tab-log}
\end{table}

\begin{figure}
 \includegraphics[width=80mm]{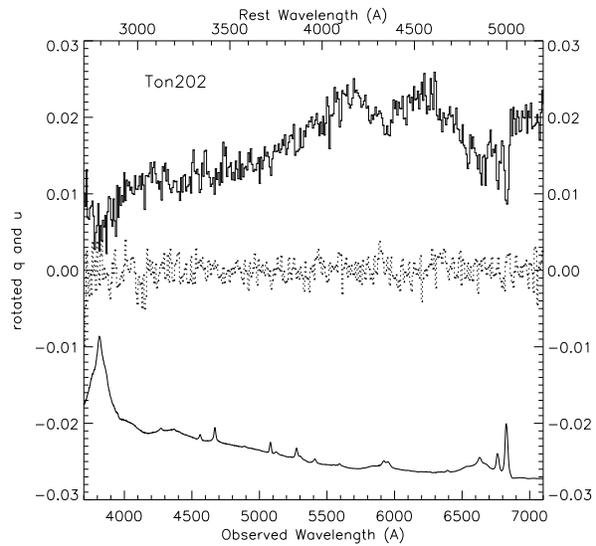}

 \caption{Rotated normalized Stokes parameters $q$ and $u$ for Ton 202 
 binned with 4 pixels ($\sim9$\AA). The bottom spectrum is the scaled
 Stokes $I$ (total flux in $F_{\lambda}$) for reference.}

 \label{Ton202_qu}
\end{figure}

\begin{figure}
 \includegraphics[width=80mm]{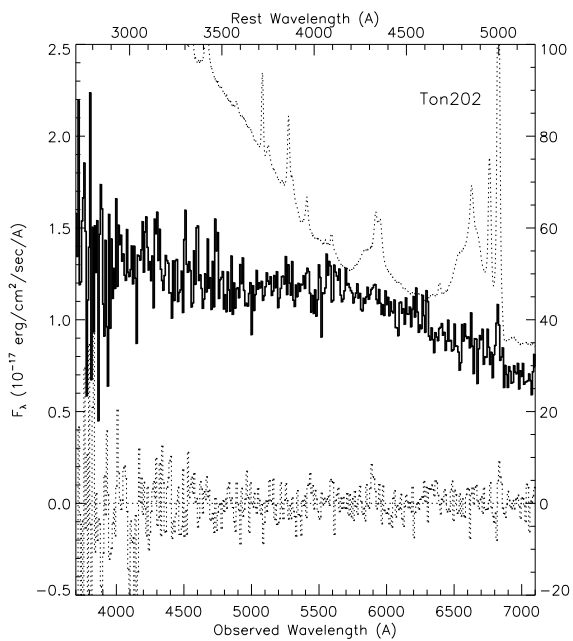}

 \caption{Rotated, unnormalized Stokes parameters $Q$ in solid line,
 essentially corresponding to $P \times F_{\lambda}$, and $U$ in dotted
 line at the bottom (flux scale is on the left axis in units of
 $10^{-17}$ erg cm$^{-2}$ sec$^{-1}$ \AA$^{-1}$) for Ton 202 binned with
 4 pixels ($\sim9$\AA), along with the scaled Stokes $I$ in dotted line
 at the top (flux scale is on the right axis also in units of $10^{-17}$
 erg cm$^{-2}$ sec$^{-1}$ \AA$^{-1}$).}

 \label{Ton202_PF}
\end{figure}

\section{Observations}\label{sec-obs}

We have observed two quasars listed in Table~\ref{tab-log} on May 8,
2002 (UT) with the Low Resolution Imaging Spectrograph (LRIS;
\citealt{Ok95}) on the Keck-I telescope using its polarimetry module.
These quasars are (1) polarized but known to have no emission lines in
the polarized flux spectrum of limited S/N in the literature (2) in the
redshift range of $z \sim 0.3$ to have the Balmer edge in the observed
frame fall on the appropriate wavelength range for the instrument
sensitivity and sufficiently far from the atmospheric cutoff.  We have
used the blue-side CCD with grism 300/5000 at a dispersion of 2.2\AA\
per pixel. The slit width was $1.''5$ and the spectral resolution was
$\sim 15$\AA.  The data were obtained with the slit approximately along
the parallactic angle. The spatial sampling was $0.''215$ per pixel.

The data were reduced in the standard manner: the bias level was
subtracted using the overscan region, frames were flat-fielded using
dome flat exposures, and obvious cosmic-ray hits were removed and
interpolated using neighboring pixels.  Then the polarimetry data
reduction was done following \citet{MRG88}.  We lost the first hours of
the night with clouds, and some of the subsequent observations might
have been through cirrus, which may have caused a little guide error. We
have checked the balance factor $\omega$ spectra \citep{MRG88} and
object positions for each pair of two frames taken with the waveplate
rotated by 45\degr, and also checked the consistency of $q$ and $u$
between different pairs.  We have excluded the pairs which had an
obvious guide problem and/or whose $\omega$ spectra had a strong
curvature over the wavelengths. The sets in which the $q$ or $u$ is
systematically different from other sets were found only in these
excluded frames.  From the 4 sets of 4 frames (4 waveplate positions)
left for Ton 202 and 2 sets left for 4C37.43 (listed in Table
\ref{tab-log}), the normalized $q$ and $u$ were obtained.  The Stokes
$I$ was obtained from the same two sets for 4C37.43, but for Ton 202, it
was obtained from three sets, excluding one which showed a slight
curvature in the $\omega$ spectrum.  The Stokes $I$'s were
flux-calibrated using the standard star BD+33d2642.

Due to a problem at the end of the night, we lost the time for the
polarization standard star observation. Therefore, there is an
uncertainty of a constant in our measurement of the position angle (PA)
of polarization. But this constant was roughly determined based on the
published PAs of our targets, and the uncertainty is probably $\sim
5$\degr\ from the probable variability of the PAs of the targets: the
observed PAs in the literature are within $\sim 5$\degr\ of 70\degr\ and
105\degr\ for Ton 202 and 4C37.43, respectively, except for one occasion
in each target (\citealt{Be90}, \citealt{SS00}; see section
\ref{sec-dis-var}).

The unpolarized standard star was also BD+33d2642. This star is reported
to be slightly polarized with $P\sim0.2$\% \citep*{SEL92}.  We also find
this level of polarization but at PA of $\sim$100\degr\ as opposed to
$\sim$10\degr\ reported by \citet{SEL92}.  We believe that the PA in
Schmidt et al. is in error by 90\degr\ for the following reasons.  (1) A
few measurements made by G. Schmidt et al. since their paper have
actually showed the PA to be 95\degr (G. Schmidt, pc, 2003).  (2) The
overall tendency of interstellar polarization around the star is at
around PA 100\degr. (3) We looked at another data set of the same star
taken in Jan 1998 at Keck (given to us by A. Barth) and found the
polarization consistent with ours. (4) Finally, another Keck observation
by Cimatti et al shows 100\degr\ (A. Cimatti, pc, 2002).  Our $P$
measurement shows a Serkowski-like curve, and matches well with Schmidt
et al's multi-band $P$ measurement, if rotated by 90\degr. Based on the
difference between these data sets, the instrumental polarization is
expected to be less than $0.05-0.1$\%.  This is also supported by the
comparison of two sets of our Ton202 observation which were taken in
different slit directions (see also section \ref{sec-dis-var}).

At the red end of the spectrum $\lambda \ga 6600$\AA, the spatial
profile degrades (worse in the observation of 4C37.43, possibly due to
larger flexure in the focus direction in LRIS from the lower attitude of
the telescope).  Therefore we have used rather large extraction windows
to accommodate it, $3.''7$ for Ton 202 and $6.''7$ for 4C37.43, although
there might still be a slight uncertainty in the P measurement at the
reddest end.

\begin{figure}
 \includegraphics[width=80mm]{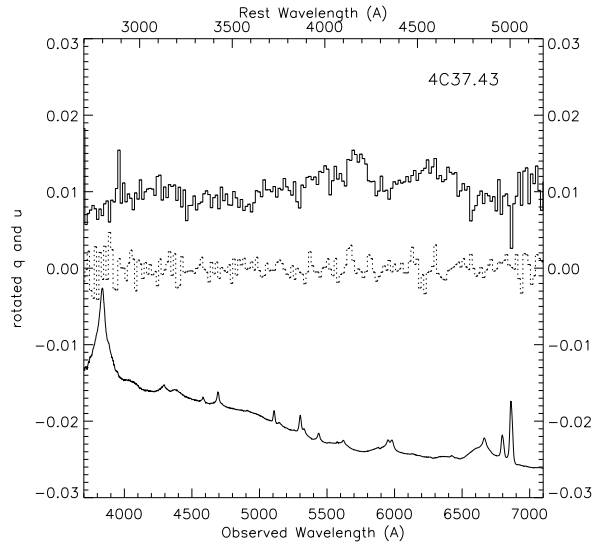}
 \caption{The same as Fig.\ref{Ton202_qu} but for 4C37.43 with 8 pixel
 bin ($\sim 18$\AA).}
 \label{4C37.43_qu}
\end{figure}

\begin{figure}
 \includegraphics[width=80mm]{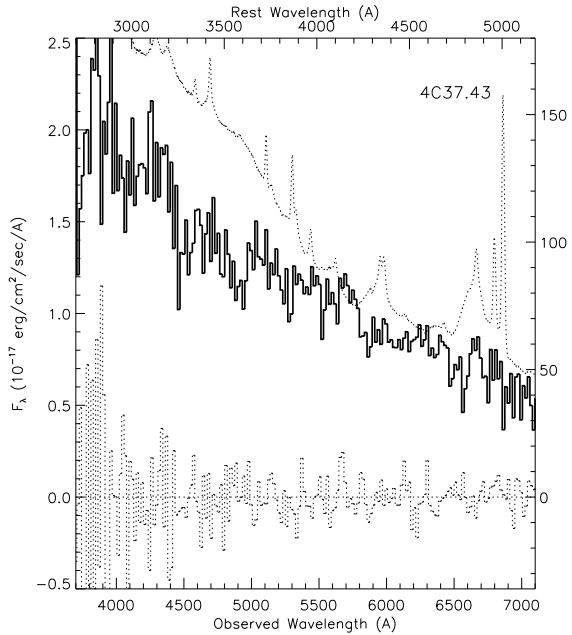}
 \caption{The same as Fig.\ref{Ton202_PF} but for 4C37.43 with 8 pixel
 bin ($\sim 18$\AA).}
 \label{4C37.43_PF}
\end{figure}

\section{Results}

\subsection{Ton 202}

Figure~\ref{Ton202_qu} shows the normalized Stokes parameters $q$ and
$u$ for the quasar Ton 202 at $z=0.366$, along with the total flux
$F_{\lambda}$ at the bottom for reference.  The reference axis of $q$
and $u$ has been rotated to be at the position angle (PA) of the
object's polarization, which is essentially wavelength
independent. Figure~\ref{Ton202_PF} shows unnormalized Stokes $Q$ in
thick lines, which is essentially the polarized flux spectrum $P \times
F_{\lambda}$, and $U$ in dotted lines at the bottom.  The dotted
spectrum at the top is the total flux $F_{\lambda}$ scaled to roughly
match the polarized flux at the red side.  The emission lines, both
broad and narrow ones, are essentially all absent in the polarized flux
to a high S/N, corresponding to the clear decrease of polarization at the
emission line wavelengths (see Fig.\ref{Ton202_qu}).

From Figure~\ref{Ton202_PF}, the slope of the polarized flux at the red
side is roughly the same as that of the total flux. However, we find
that the slope of the polarized flux changes at the bluer side of
$\sim$4000\AA\ in the rest frame (note the rest wavelength scale at the
top of Figures), just where the small blue bump starts in the total
flux. There might also be a possible up-turn at the bluer side of
$\sim$3600\AA\ in the rest frame.  To illustrate the slope change
quantitatively, if we measure the slope at the blue side (2800 -
4000\AA\ in the rest frame) and the red side (4000 - 4900\AA) by fitting
a broken power law, we obtain $\alpha = -1.73 \pm 0.07$ and $-0.54 \pm
0.08$, respectively, where $F_{\nu} \propto \nu^{\alpha}$ (we have
excluded the red end of the spectrum considering the possible
uncertainty from the spatial profile degradation; see
section~\ref{sec-obs}).  Note that this spectral slope change is too
sharp to be explained by dust reddening: we tried adopting available
reddening laws (e.g. \citealt{Ca89}) on a power-law continuum, but no
reasonable fit was obtained.

We identify this slope change (and the possible secondary up-turn)
to be a Balmer edge feature, seen in absorption in the shorter
wavelength side. It certainly does not look like an ``edge'' seen in the
continuum opacity. This is conceivable, since the real edge should
consist both of the continuum edge and high order Balmer lines, so that
the combined effect may well be that the edge feature starts rather at
around 4000\AA, instead of 3646\AA.  In addition, there may be
absorption by metal lines, resembling the 4000\AA\ break in an old
stellar population. Recall too that there may be a secondary dip/up-turn
at $\sim3600$\AA\ near where the actual edge is expected.

However, admittedly, the edge and absorption lines probably need to be
broadened in order to be consistent with the observed spectral shape of
the polarized flux.  This could be attributed to a Doppler smearing due
to a high velocity dispersion expected in the deep potential well at the
nucleus (though in terms of accretion disc models, the smearing effect
is smaller for the Balmer edge than for the Lyman edge, as discussed in
section~\ref{sec-intro}).  Without referring to any particular model,
however, the shape of the feature and the wavelength of the slope change
strongly suggest that the feature is most likely to be a Balmer edge
absorption feature.

\subsection{4C37.43}

The data for 4C37.43 ($z=0.371$) are shown in Figures~\ref{4C37.43_qu} and
\ref{4C37.43_PF} in the same manner as for Ton 202. The data are of
lower S/N compared to those for Ton 202, but there might be a weak broad
H$\beta$ emission line in the polarized flux in this object.  This means
that there might also be some contribution from the small blue bump in
the polarized flux of this object, making our scraping method
incomplete.  Apparently, there does not seem to be a clear slope change
in the polarized flux, but the plot in $\nu F_{\nu}$
(Fig.\ref{4C37.43_nFn}) might suggest a very slight slope change again
around 4000\AA. This is still unclear in the slope measurements with a
broken power law: $\alpha = -0.17 \pm 0.11$ where $F_{\nu} \propto
\nu^{\alpha}$ in the blue side (2800-4000\AA\ in the rest frame) and
$\alpha = +0.06 \pm 0.23$ in the red side (4000-4700\AA, excluding the
possible broad line component and the red-end portion).

\begin{figure}
 \includegraphics[width=80mm]{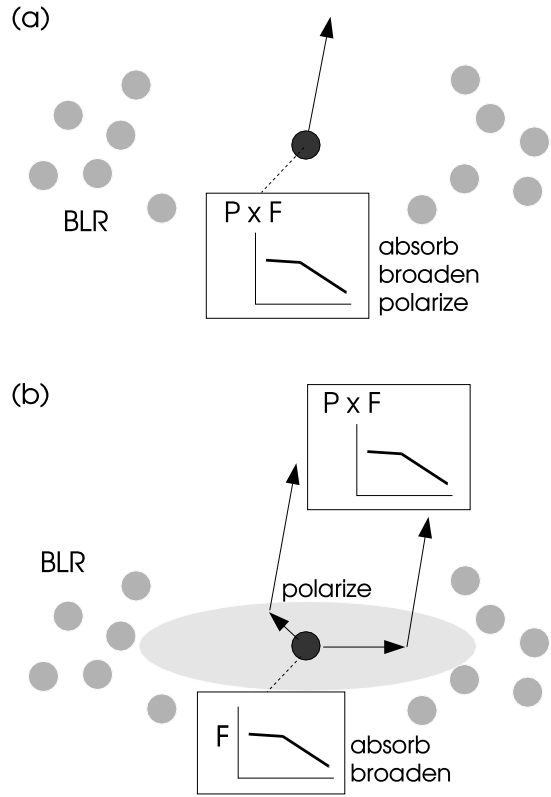}
 \caption{Schematic drawings of the possible geometric
 configurations. (a) The polarized flux is coming from the Big Blue Bump
 emitter at the centre. (b) The polarized flux is formed via scattering
 in the oblate electron scattering region surrounding the Big Blue Bump
 emitter. }
 \label{polmodel1}
\end{figure}

\section{Discussion}

\subsection{The nature of the Big Blue Bump emission}\label{sec-dis-bbb}

As we have pointed out above, there are essentially no broad (and narrow)
emission lines in the polarized flux of Ton 202, and the polarization is
confined to the continuum.  This means that the polarization
seen here is very different from the polarization seen in many Seyfert 2
galaxies. In the latter, the continuum emission {\it and} the broad line
emission are {\it both} scattered in the region exterior to the BLR and
coming into our line of sight, so that both the continuum and broad
lines are present in the polarized flux. In these Seyfert 2's, the
position angle of polarization is perpendicular to the radio jet
structure. These two facts have lead to the thought that the scattering
region is located along the jet axis outside the nucleus and the BLR.

In contrast, in the case of Ton 202 where the broad lines are shown to
be essentially unpolarized (and in other quasars shown in Schmidt and
Smith 2000, though with a limited S/N), the region which is causing the
polarization should be {\it interior} to the BLR (unless the polarized
light is a completely different component from the BBB emission; see the
discussion below in section \ref{sec-dis-oth}).  Unlike the cases of the
Seyfert 2's above, we still do not have a good understanding of the
polarization mechanism in these quasars.  One of the main difficulties
in disc atmosphere calculations is that disk atmospheres generally
produce polarization perpendicular to the disk axis, while the position
angle of polarization is observed to be {\it parallel} to the jet axis.
(The difference between the optical polarization PA and the radio jet
axis is 15\degr\ and 9\degr\ for Ton 202 and 4C37.43, respectively, from
\citealt{SS00}; given the polarization PA variability and the
imperfectly measurable radio axis PA, the differences from zero are not
necessarily real.)  However, the non-existence of exact models for the
parallel polarization does not affect our conclusion that the polarizing
region (the region causing the polarization) should be interior to the
BLR (see more below for the discussion of the size).

Geometrically, we can first think of two cases.  (1) The source of the
BBB emission and the polarizing region are essentially the same --- the
Balmer-edge absorption, its broadening, and polarization are occurring
within the BBB emitter (see Fig.\ref{polmodel1}a). One such an example
is a disk atmosphere (aside from the inconsistency in polarization
direction).  The Balmer edge feature in the polarized flux can mean
either that the total flux of the BBB has an edge feature with a
wavelength-independent polarization, or the total flux is rather
featureless but the polarization is low at shortward of the edge.  In
either case, the edge feature in the polarized flux is simply intrinsic
to the BBB emission in the sense that the feature is formed within the
BBB emitter.  (2) The polarizing region is located somewhere between the
BBB emitter and the BLR.  One conceivable case is an optically-thin
oblate electron scattering region surrounding an unpolarized BBB emitter
(see Fig.\ref{polmodel1}b). The region would be optically thin (see
discussion below), oblate and perpendicular to the jet axis in order to
produce the polarization parallel to the axis (see \citealt{BM77} for
polarization calculation).  We would not consider dust scattering since
the region is essentially within the dust sublimation radius.  The
intrinsic BBB polarization might have been completely wiped out by
Faraday depolarization.  In this optically-thin scattering case, the
polarized flux spectrum is simply proportional to the incident BBB
emission, so that the edge feature should have been formed, again,
within the BBB emitter.

Therefore, in both cases above, the Balmer-edge absorption feature
identified in the polarized flux is intrinsic to the BBB emission. Thus
the feature indicates that the BBB emission is indeed thermal. The fact
that the edge feature is seen in absorption in the shorter wavelength
side indicates that the emission is optically thick. This is our first,
primary conclusion.

The possible weak broad line component in the polarized flux of 4C37.43
might suggest that the size of the polarizing region in this object is
not much smaller than the size of the BLR, which might be the case {\it
in general}.  In this sense, the case (2) above may be preferred rather
than case (1). The radius of the oblate electron scattering region might
be extended as large as the inner radius of the BLR.

Many Seyfert 1 galaxies also show a few \% level polarization in
continuum (e.g. \citealt{GM94}; \citealt{Sm02}). When the radio
structure is available, the radio axis is parallel to the optical
polarization direction in the majority of the cases
\citep{An83,An02,Sm02}.  In contrast to our two quasars, broad lines are
very clearly polarized in these Seyfert 1 galaxies.  However, the broad
line polarization integrated over the line profile is generally at a
different PA than the continuum, and the PAs often show some systematic
rotation within the line wavelengths \citep{GM94,Sm02}.  This might
suggest that the polarizing region in these Seyfert 1s is not much
larger than the BLR, and it may be closely related to the oblate
scattering region discussed above.

\subsection{Other possibilities}\label{sec-dis-oth}

\begin{figure}
 \includegraphics[width=80mm]{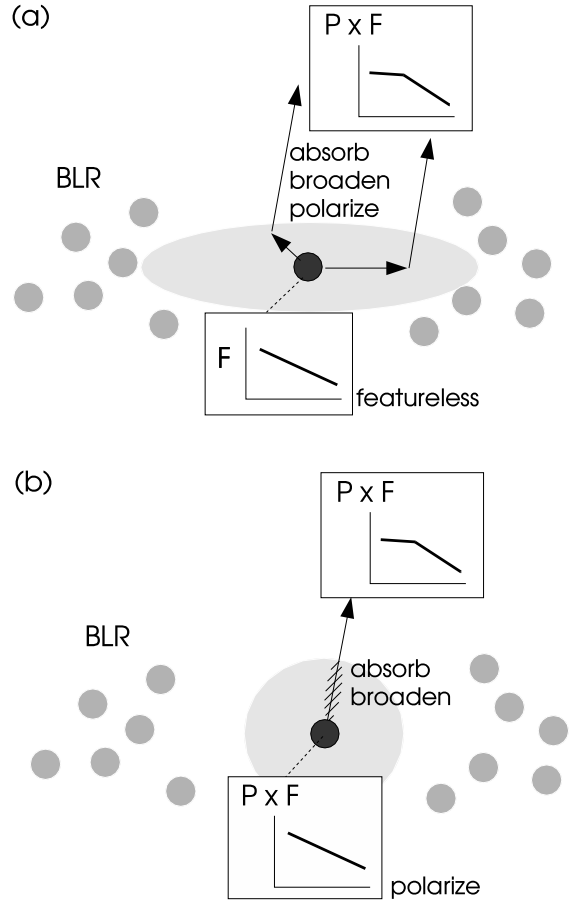}

 \caption{Schematic drawings for other possibilities. (a) The
 Balmer-edge absorption feature is formed within the
 scattering/absorbing region surrounding the Big Blue Bump emitter. (b)
 A broadened Balmer-edge feature is imprinted in a foreground region on
 an incident featureless polarized flux.  \label{polmodel2}}

\end{figure}

The two cases above are probably the simplest cases for producing the
observed polarization.  However, we cannot rule out some cases where the
Balmer edge feature seen in the polarized flux is still not intrinsic to
the BBB emission.  There might be a case where the scattering medium in
case (2) above is also imprinting the absorption feature on the
polarized flux spectrum (while the incident BBB spectrum is featureless;
see Fig.\ref{polmodel2}a). In this case, however, we need to have an
absorption optical thickness of order unity in the Balmer continuum
along the size of this region.  The broadening of the absorption feature
also need to occur in this scattering/absorbing medium. This would
require a rather large velocity dispersion, having to put this region,
for example, in a rather deep potential well.  Having too a significant
absorption optical thickness, especially in a deep potential well, is
rather equivalent to having the whole scattering region as the BBB
emitter itself, so that the distinction becomes largely semantic.  This
region might also produce a significant soft X-ray absorption.  The
broadening might be due to the scattering by hot electrons, but in this
case, a significant alteration of the X-ray Fe K$\alpha$ line might be
expected (but see \citealt{CZ91}).

Alternatively, the Balmer absorption feature could be imprinted in the
featureless polarized flux by some foreground region
(Fig.\ref{polmodel2}b). The background source of a featureless polarized
flux could be the BBB emitter itself or a completely different source
(in this case, a high polarization is required for this secondary
source), assuming that the polarized flux is somehow produced by
e.g. synchrotron process or electron scattering within that source.
This case would require some particular velocity structure in this
foreground medium along the line of sight, to produce the apparent slope
change.  Also the absorption feature would be more significant in the
Lyman edge region in this case, since the Lyman transition is a resonant
one. Such Lyman edge absorption is not seen in quasars, though the UV
data for Ton 202 itself are quite noisy (\citealt{La93}; the UV spectrum
of 4C37.43 looks featureless at the Lyman edge).  Note that, conversely,
the slope changes at the Lyman edge region seen in some cases, as
discussed in section \ref{sec-intro}, could be a foreground absorption,
but this is less likely to be the case for the (non-resonant) Balmer
edge region.

\citet{SS00} preferred synchrotron emission for the explanation of the
optical polarization of the lobe-dominant quasars which show the
alignment of the polarization PA and radio axis, including our two
targets here.  The polarized flux of Ton 202 in their observation seemed
featureless, while it was not in our observation, which could be due to
a limited S/N in their observation but also a variability (see the next
section).  If synchrotron emission is the origin of the polarization,
the observed Balmer-edge feature is thought to be imprinted in the
polarized flux somewhere along the line of sight, which is what we have
just discussed. While we cannot rule this out, we need to consider the
critical implications described above.

More generally, we note that synchrotron emission from the misdirected
blazar core components, which is favored by \citet{SS00}, does not seem
to be enough to explain the observed optical polarized flux level of
lobe-dominant quasars. For our two objects specifically, the ratio of
the radio core flux to the optical {\it polarized} flux in $F_{\nu}$ is
$\sim 2 \times 10^3$ (see Table 1 in \citealt{SS00}).  Thus the ratio of
the radio core flux to the putative optical synchrotron flux is $\sim 2
\times 10^3 \times P'$, where $P'$ is the fractional polarization of the
optical synchrotron emission.  Since the radio cores of lobe-dominant
quasars typically become optically thin by 30GHz \citep{An90}, the
expected ratios would be more like $\sim 2 \times 10^4$ (for $F_{\nu}
\propto \nu^{-1}$).  Also, it is known that the optical polarization
distribution for lobe-dominant radio loud quasars is similar to that for
radio quiet quasars \citep{St84}.

\subsection{Variability}\label{sec-dis-var}

Our polarization measurement of 4C37.43 looks rather consistent with
that of \citet{SS00} in the red part of the spectrum. But in the bluer
part, the $P$ in our measurement is slightly higher by $\sim 0.1-0.2$\%,
and the total flux looks also slightly bluer, leading to a bluer
polarized flux color ($\alpha = -0.11 \pm 0.07$ for $\lambda_{\rm rest}
= 2800-4700$\AA\ in our data, whereas $\alpha = -0.8 \pm 0.2$ in
\citealt{SS00}).  Compared with the data for Ton 202 in \citet{SS00} and
Antonucci (1988; data of Miller and Goodrich), the measured polarization
in our data is slightly higher especially at 4000-4800\AA\ in the rest
frame of the quasar. The difference in $P$ is up to 0.5\%, which is much
larger than the instrumental polarization level discussed in section
\ref{sec-obs}.  In the unlikely case that the identified edge-like
feature in the polarized flux is solely due to the instrumental
polarization, the $q$ and $u$ spectrum in the sky coordinates should
look different in the data taken at different instrument angles (see
Table~\ref{tab-log}), if the instrumental polarization stays the same in
the instrumental coordinates.  Since the instrument angles are different
by 30\degr, the instrumental polarization component should rotate by
60\degr\ in the $q$-$u$ plane of the sky coordinates, which should
obviously show up in the $u$ spectrum.  This is not the case, and the
polarized flux spectra from two different instrument angles show
essentially the same feature.

Therefore, the polarization of Ton 202, and probably also 4C37.43, seems
to have varied slightly over these several years (from 1996-1999 to
2002). This is actually conceivable, since the polarization arises from
a region smaller than (or in general, at most comparable to) the BLR, as
we argued above. As noted in section \ref{sec-obs}, the polarimetric
data of \citet{Be90} and \citet{SS00} which span from 1978 to 1999 have
some evidence of variability for both of the objects (see section 4.2 of
\citealt{SS00}). Similar probable variations of the polarization have
been reported for broad line radio galaxies (e.g. \citealt{An84}).  The
polarization of the continuum and broad lines in Seyfert 1 galaxies,
discussed in section \ref{sec-dis-bbb}, are seen to vary on similar time
scales (e.g. \citealt{Yo99}, \citealt{Sm02}).  We plan to follow up our
objects in an upcoming observing run.

\subsection{Behavior in the shorter wavelength region and other targets}

\begin{figure}
 \includegraphics[width=80mm]{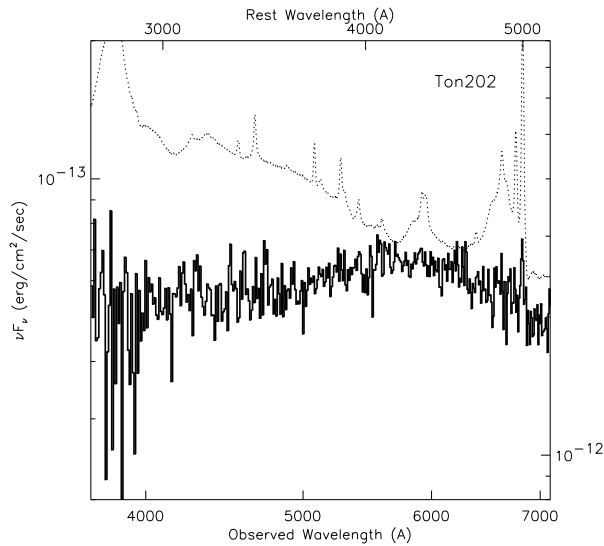}

 \caption{ $\nu F_{\nu}$ (or $\lambda F_{\lambda}$) plot for Ton 202,
 with both axes on log scale. The solid line is for the polarized flux,
 i.e. $\lambda$ multiplied by rotated unnormalized Stokes $Q$ parameter,
 essentially corresponding $\lambda P F_{\lambda}$ (the scale is on the
 left axis in units of erg cm$^{-2}$ sec$^{-1}$). The dotted line is the
 scaled $\lambda F_{\lambda}$ (the scale is on the right axis also in
 units of erg cm$^{-2}$ sec$^{-1}$).}

 \label{Ton202_nFn}
\end{figure}

\begin{figure}
 \includegraphics[width=80mm]{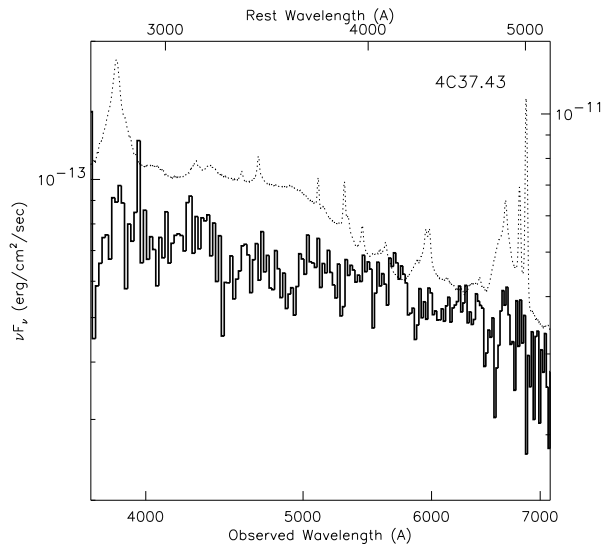}
 \caption{The same as Fig.{\ref{Ton202_nFn}}, but for 4C37.43.}
 \label{4C37.43_nFn}
\end{figure}

Figures \ref{Ton202_nFn} and \ref{4C37.43_nFn} shows the same data as in
Figure~\ref{Ton202_PF} and \ref{4C37.43_PF} but in $\nu F_{\nu}$ instead
of $F_{\lambda}$, with both axes on log scales.  For Ton 202, $\nu
F_{\nu}$ has a peak in the wavelength around the Balmer edge feature at
least locally.  If the total flux has the same spectral shape as the
polarized flux, as in the case of polarization being produced in the
surrounding electron scattering region or in the case of a
wavelength-independent polarization of the BBB emitter, this implies
that the energy output from the small blue bump is much larger than
previously thought. Also, in order for the UV continuum to ionize the
surrounding region, the BBB emission should turn up as it goes into the
shorter wavelength region under the small blue bump. The possible
up-turn observed in the polarized flux is still uncertain and will be
followed up with a new blue sensitive CCD at the Keck. Targets with
higher redshifts would of course be better for the investigation at
shorter wavelengths.

As for statistics and other targets, we have observed and presented in
this paper two quasars among the ones which previously appeared to show
polarization confined to continuum (no polarized broad line) in a
limited S/N.  One of them, Ton 202, is confirmed to have no polarized
broad line in our high S/N observation, and has shown a clear slope
change in the polarized flux. The result for the other quasar, 4C37.43,
is not still conclusive.  In \citet{SS00}, there are 5 more quasars with
continuum-confined polarization in a limited S/N.  Their polarized flux
should be observed with a very high S/N so that the broad lines and the
Balmer edge wavelength region can be closely examined. We have further
observed some of them at the VLT, which will be published elsewhere.

\subsection{Broadening and Accretion Disk Models}

As we have pointed out, even though the high order Balmer absorption
lines, and possibly metal lines, are expected to blur the edge
substantially, the observed edge feature suggests a broadening probably
by a high velocity dispersion of the gas around the nucleus.  In this
case, very crudely, the velocity dispersion needed is $\Delta \lambda /
\lambda \sim 400/3600 \sim 0.1c$, corresponding to $\sim 100 R_g$ ($R_g
= G M /c^2$ for a black hole mass $M$) for a Keplerian rotation.

In terms of accretion disc models for quasars, this radius is broadly
conceivable for a Balmer edge emitting radius.  The broadening depends
on the inclination of the disc to the line of sight: our quasars are not
expected to be so close to face-on, based on the observed core-to-lobe
radio flux ratios $R$ ($log R = -0.40$ and $-0.71$ for Ton 202 and
4C37.43, respectively, from \citealt{Wi92}, though this is a very crude
indicator). Then, the Doppler broadening would be effective.  The
preliminary calculation using the model of \citet{Hu00} without any
absorption lines show that the emission from the disc atmosphere can
produce at least very crudely such a smeared Balmer edge feature. Note
again that the high order Balmer absorption lines can be crucial to the
spectral shape at the edge region. An appropriate model is under
development.  We stress, however, that our conclusion of an
optically-thick thermal BBB emitter does not involve any particular
model of the nucleus.

\section{Conclusions}

We have taken high S/N polarized flux spectra of two quasars with LRISp
on the Keck telescope in order to investigate the behavior around the
Balmer edge.  One of them, Ton 202, shows a slope change at around
4000\AA. There might also be a possible up-turn at 3600\AA\ in the same
polarized flux spectrum. We identify this to be a Balmer edge feature,
seen in absorption in the shorter wavelength side.  There are
essentially no broad emission lines in the polarized flux spectrum of
Ton 202. This indicates that the polarized flux arises interior to the
BLR, showing the spectrum with all the contaminating emissions from the
BLR and outer regions scraped off and removed.  Therefore, the polarized
flux spectrum shape is likely to be intrinsic to the Big Blue Bump
emission. In this case, the identified Balmer edge feature indicates
that the Big Blue Bump emission is thermal and optically thick.

The polarized flux shape would be intrinsic to the BBB emission in two
general cases: (1) the Balmer-edge absorption, its broadening, and
polarization are occurring within the BBB emitter itself or (2) the
broadened Balmer-edge absorption feature is formed within the BBB
emitter, and some part of the emission is subsequently scattered (thus
getting polarized) by electrons in an optically-thin region surrounding
the BBB emitter.  This region would be oblate in shape and perpendicular
to the jet axis in order to produce polarization parallel to the axis.

In another quasar, 4C37.43, there might be a very weak broad component
of H$\beta$ line in the polarized flux.  This could imply that the
polarization arises from scales not much smaller than the BLR. This
might be the case in general, so that the case (2) above may be
preferred and the size of the postulated oblate scattering region might
generally be comparable to the innermost radius of the BLR.

However, we cannot rule out some cases where the Balmer-edge feature is
not intrinsic to the BBB emission.  The Balmer-edge absorption feature
might be formed within the scattering region surrounding the featureless
BBB emitter, along with the polarization. However in this case, the
scattering/absorbing region should be rather optically thick and in a
rather deep potential well, which might make this case almost equivalent
to the case (1) above.  It might also be possible to postulate (possibly
independent) compact featureless polarized flux emitter and have the
absorption and broadening in a foreground region, but this would require
a particular velocity structure, and the absorption at the Lyman edge
would be implausibly large.

\section*{Acknowledgments}

The authors thank A. Barth, D. Leonard, A. Cimatti, G. Schmidt for
helping us to clarify the polarization measurements on the null standard
star, E. Agol for providing calibration data, R. Goodrich for various
information on the Keck polarimetry, and A. Laor for useful comments on
the manuscript of the paper. The work by RA was supported in part by NSF
grant AST-0098719, and the work by OB was supported in part by NASA
grant NAG5-7075.


\label{lastpage}

\end{document}